\documentclass[preprint]{JHEP3} 

\usepackage{graphicx}
\usepackage[tight]{subfigure}
\usepackage{amsmath}
\usepackage{multirow}

\newcommand\fverb{\setbox\pippobox=\hbox\bgroup\verb}
\newcommand\fverbdo{\egroup\medskip\noindent%
                        \fbox{\unhbox\pippobox}\ }
\newcommand\fverbit{\egroup\item[\fbox{\unhbox\pippobox}]}

\newcommand{\de}[2]{(\delta_{#1}^d)_{#2}}

\newcommand{\colorparbox}[3]{\colorbox{#1}{\parbox{#2}{#3}}}
\newcommand{\colorwideparbox}[2]{\colorparbox{#1}{\textwidth-5mm}{#2}}

\newcommand{\chapter}[1]{\color{fgSection}\colorwideparbox{bgSection}{%
    \centering\bfseries\boldmath#1}}

\newcommand{\g}[3]{\Gamma_#1^{#2#3}}

\newcommand{\GeV}{\ \mathrm{GeV}}

\newcommand{\tb}{\tan\!\beta}
\newcommand{\bsg}{B \rightarrow X_s \gamma}
\newcommand{\dms}{\Delta M_s}
\newcommand{\bsbsbar}{$B_s$--$\overline{B_s}$}
\newcommand{\ACPbsgam}{A_{CP}^{b \rightarrow s \gamma}}

\newcommand{\SphiK}{S_{CP}^{\phi K}}
\newcommand{\SetapK}{S_{CP}^{\eta' K}}
\newcommand{\SKstargam}{S_{CP}^{K^* \gamma}}
\newcommand{\phis}{\phi_s}

\newcommand{\deltad}{\delta^d}
\newcommand{\ded}[2]{(\deltad_{#1})_{#2}}

\newbox\pippobox

\title{Implications of the measurements of
{\boldmath\bsbsbar} mixing on SUSY  models}

\author{P. Ko  \\
        School of Physics, KIAS, Seoul 130-722,  Korea \\
        E-mail: \email{pko@kias.re.kr}}

\author{Jae-hyeon Park  \\
        INFN, Sezione di Padova, via F Marzolo 8, I--35131, Padova, Italy\\
        E-mail: \email{jae-hyeon.park@pd.infn.it}}

\received{}            
\revised{}
\accepted{}            

\preprint{}

\abstract{%
We derive constraints on the mass insertion parameters from the recent
measurements of \bsbsbar\ mixing, and discuss their
implications on SUSY breaking mediation mechanisms and SUSY flavor models.
Some SUSY flavor models are already excluded or disfavored by
\bsbsbar\ mixing.
We also discuss how to test the SM and SUSY models in the future
experiments, by studying other CP violating observables related to
$b\rightarrow s$ transition, such as the time-dependent
CP asymmetry in $B_d \rightarrow \phi K_S$ and the direct CP asymmetry in
$B\rightarrow X_s \gamma$.}

\begin{document}

\section{Introduction}

Within the Standard Model (SM) with three families, there is a unique
source of flavor and CP violation in the quark sector, which is the
renowned Cabbibo-Kobayashi-Maskawa (CKM) mixing matrix \cite{km}. 
The CKM paradigm has long been tested in the $K, D$ and $B$ meson 
systems during the last decades. As of now, this picture has been well
confirmed to describe basically all the data related with flavor
and CP violation in the quark sector, modulo some theoretical and
experimental uncertainties. Experimental uncertainties will be
decreased as more data are taken at B factories, whereas
theoretical uncertainties will be under better control when more
results come from unquenched lattice QCD simulations on various
nonperturbative parameters that are relevant to CKM analysis.

For many years, one of the important ingredients in the CKM
phenomenology was still missing, namely $\Delta M_s$ from
\bsbsbar\ mixing. Recently, however,  $\Delta M_s$ was
measured by both D\O\ and CDF Collaborations at the Tevatron:
\begin{equation}
17 ~{\rm ps}^{-1}  < \Delta M_s < 21 ~{\rm ps}^{-1} ~~~ \text{(D\O) }
\cite{d0} ,
\end{equation}
\begin{equation}
\Delta M_s  =  ( 17.77 \pm 0.10 \pm 0.07 ) ~{\rm
ps}^{-1}~~~ \text{(CDF) } \cite{Abulencia:2006ze} .
\end{equation}
One can use the measured value of $\Delta M_d / \Delta M_s$ to
determine $| V_{td} / V_{ts} |$ within the SM \cite{Abulencia:2006ze}:
\begin{equation}
| V_{td} / V_{ts} |  = 0.2060 \pm 0.0007\ (\dms) ^{+0.0081}_{-0.0060}~~
(\Delta M_d + \text{theor}).
\end{equation}
This result is consistent with another independent determination
of $| V_{td} / V_{ts} |$ from the Belle measurement of a  radiative 
decay $B\rightarrow X_d \gamma$ \cite{bdgamma}:
\begin{equation}
| V_{td} / V_{ts} |  = 0.199^{+0.026}_{-0.025} ({\rm exp})^{+
0.018}_{-0.015} ({ \rm theor}) .
\end{equation}
Excellent agreement of these two independent measurements
constitutes another firm test of the CKM paradigm for flavor and
CP violation in the SM \cite{ckmfit,hfag}, and puts strong constraints 
on various new physics scenarios. There are model independent analyses of 
$\Delta M_s$ measurements  on general new physics 
\cite{nir,lenz}, 
as well as analyses within supersymmetric (SUSY) models \cite{others,Foster:2006ze} 
and others \cite{others2}. 
Due to these data on $\Delta M_s$, the CKM paradigm is more
constrained than before, and  
there may be even a slight hint for new physics beyond the SM 
(see Ref.~\cite{lenz}, for example).

Within the SM, \bsbsbar\ mixing is dominated by
$t$--$W$ loop, and the \bsbsbar\ mixing phase is
suppressed by $\lambda^2$ \cite{sm_review}.   
Due to its small theoretical uncertainty, 
observation of a nonzero discrepancy in the phase of \bsbsbar\
mixing 
would be an unambiguous signal of
new physics beyond the SM in $b\rightarrow s$ transition
\cite{Fleischer:2008uj}.
Such new physics effects, if any, may appear in other observables in
the $B_{(d,s)}$ meson systems, e.g., $B_d \rightarrow \phi K_S$ or
$B \rightarrow X_s \gamma$.

The two collaborations also reported results on the phase of 
$B_s - \overline{B_s}$  mixing from the time dependent CP asymmetry in
$B_s \rightarrow J/\psi \phi$ and
the charge asymmetry
(or CP violation in the mixing) in the $B_s$ system:
\begin{equation}
A_{\rm SL}
\equiv {N(B_sB_s) - N(\overline{B_s}\overline{B_s}) \over
  N(B_sB_s) + N(\overline{B_s}\overline{B_s})}
\approx
{\rm Im} \left( { \Gamma_{12} \approx \Gamma_{12}^{\rm SM} \over
M_{12}^{\rm SM} + M_{12}^{\rm SUSY} } \right) .
\end{equation}
The results are
\begin{align}
  \phis &= -0.57^{+0.24}_{-0.30}(\text{stat})^{+0.07}_{-0.02}(\text{syst})
  \quad \text{(D\O) } \cite{Abazov:2008fj} ,
  \\
  \phis &\in [-1.36,-0.24] \cup [-2.90,-1.78]
  \quad \text{(CDF) } \cite{Aaltonen:2007he} ,
\end{align}
at 68\% CL\@.
These measurements give a strong constraint on the new physics
contributions to \bsbsbar\ mixing,
both the modulus and the phase of the mixing.
In general SUSY models, this will constrain the $23$ mixing,
$( \delta_{23}^d )_{AB}$ with $A,B=L$ or $R$.

In this paper, we update our previous studies on $b\rightarrow
s$ transitions within the general SUSY models \cite{Kane:2003zi,Kane:2002sp}   
using the new data on $B_s$ mixing from D\O\ and CDF, and discuss their
implications for SUSY models. In Sec.~2, we describe the general SUSY 
models with gluino-mediated flavor/CP violation in brief, 
and how to proceed and analyze the SUSY models. 
Compared with the previous studies, we consider the $\tan\beta$
dependent constraint carefully including the double mass
insertions, which can be prominent in $B\rightarrow X_s \gamma$
for large $\tan\beta$. 
In Sec.~3, we present the constraints on 
the mass insertion parameters for several different scenarios: the
$LL$ or the $RR$ dominance case, and $LL = \pm RR$ cases. We also
mention briefly the implications of our results on $B_s
\rightarrow \mu^+ \mu^-$, which can give important informations on
$\delta$'s in the large $\tan\beta$ region. In Sec.~4, we discuss
implications of the newly derived bounds on the mass insertion
parameters on SUSY models. Most SUSY models with universal soft
scalar masses at some high energy scale or many SUSY models with
flavor symmetry groups are still consistent with our new
constraints. But some SUSY flavor models based on flavor symmetries 
and alignment of quark and squark mass matrices are shown to be in
conflict with our constraints, and thus excluded or disfavored, 
depending on $\tan\beta$. In Sec.~5, we summarize our results
and discuss the prospects in the future directions in theory and
experiments which should be taken in order to test the CKM paradigm 
and see by any chance some new physics effects lurking in 
$b\rightarrow s$ transitions.

\section{Models and analysis procedures}

\subsection{Gluino-mediated flavor violation and mass insertion approximation}

The minimal supersymmetric standard model (MSSM)
has many nice motivations such as 
resolution of fine tuning problem of Higgs mass parameter, 
gauge coupling unification,  and cold dark matter \cite{mssm}.
But SUSY, if it exists,  must be broken, and SUSY breaking 
effect is described phenomenologically by more than 100 new parameters
in the so-called soft SUSY breaking lagrangian.  
These soft SUSY breaking parameters generically violate both flavor and CP\@.
If these parameters take generic values, one ends up with excessive flavor
and CP violations which are already inconsistent with such low energy data
as $K^0$--$\overline{K^0}$ mixing, $\epsilon_K$, $B\rightarrow X_s \gamma$ 
and electron/neutron electric dipole moments (EDMs).
Therefore, there must be some mechanism which controls the structures of
flavor changing neutral currents (FCNCs)
and CP in the soft SUSY breaking terms, if weak scale SUSY has 
anything to do with Nature.
This may be achieved by means of the SUSY breaking mediation mechanism 
which is flavor blind, and/or some flavor symmetry controlling both Yukawa 
couplings and sfermion mass matrices in flavor space. 
In a different point of view, we could get a clue to these SUSY breaking 
mediation mechanisms by studying FCNC and CP in supersymmetric models.

In SUSY models, there are new contributions to \bsbsbar\
mixing from $H^{-}$--$t$, $\chi^-$--$\tilde{U}_i$
and $\tilde{D}_i$--$\tilde{g} ( \tilde{\chi}^0 ) $ in addition to the SM
$t$--$W$ loop. In generic SUSY models, the squark-gluino loop contribution 
is parametrically larger than other contributions, since it is
strong interaction. In this work, we assume that the dominant SUSY
contribution to $B_s$ mixing comes from down
squark-gluino loop diagrams. This assumption simplifies the numerical 
analysis considerably. Including effects from other SUSY
particles is straightforward, and similar analysis could be done.
A similar analysis for the $b\rightarrow d$ transition has been
performed within the mass insertion approximation \cite{kramer},
using $B_d$--$\overline{B_d}$ mixing, $A_{\rm SL}^d$ and CP
violation in $B \rightarrow X_d \gamma$ under the same
assumptions.  

Mass insertion approximation is a useful tool to present flavor 
and CP violations in the sfermion sector in generic SUSY models 
\cite{mia}.
The parameter   \( (\delta_{ij}^d)_{AB} \)
represents the  dimensionless transition strength from
$\widetilde{d}_{jB}$ to $\widetilde{d}_{iA}$ in the basis where the
fermion Yukawa couplings are diagonal (SCKM basis),
where $i,j = 1,2,3$ are generation indices and
$A,B = L,R$ are chiralities of superpartners of quarks%
\footnote{A quantitative definition of the $\delta$ parameters
will be given below.}.
If $( \delta^d_{ij} )_{AB} \sim O(1)$, there are excessive FCNC and CP
violations  with strong interaction couplings, which are clearly excluded
by the data. Therefore $\delta$'s should be small,
$\lesssim 10^{-1}$--$10^{-3}$ with upper bounds depending on
$(i,j,A,B)$, 
which is so called the SUSY FCNC/CP problem.

Current global analysis of the CKM matrix elements indicates that
any new physics around TeV scale should be flavor/CP blind to a
very good approximation. Therefore it would be nice if we can set
$\delta=0$. However, even if we set $\delta$'s to zero by hand at
one energy scale (presumably at high energy scale), nonzero
$\delta$'s are regenerated at electroweak scale due to the
renormalization group (RG) evolution, and we cannot make
$\delta$'s vanish at all scales. It is most likely that $\delta$'s
are nonvanishing at electroweak scale. Then, the relevant
questions are how large or small $\delta$ parameters are in a given
SUSY breaking scenario, and what are the observable consequences
of nonzero $\delta$'s in flavor and CP violation beyond the
effects derived from CKM matrix elements. These issues will be
addressed in the subsequent sections.

Since flavor physics and CP violation such as
$B\rightarrow X_s \gamma$, $B_s \rightarrow \mu^+ \mu^-$,
$\epsilon_K$ within SUSY models depend strongly on soft 
SUSY breaking sector which is not well understood yet, it is important 
not to make an {\it ad hoc} assumption on the soft terms.
For example, the usual assumption in the mSUGRA scenario is not well
motivated theoretically, although it seems acceptable phenomenologically
since it solves the SUSY flavor and CP problem. However such
assumptions are made for the sake of simplicity in studying
flavor physics, dark matter
and collider physics signatures within SUSY context. Sometimes, it gives
wrong intuitions, some examples of which can be found in Ref.s
~\cite{flavor_dm}.

In the following, we at first consider $\delta$'s as free parameters
at the electroweak scale, and derive phenomenological constraints on these
parameters, including $B \rightarrow X_s \gamma$ and the newly
measured \bsbsbar\ mixing. Then we estimate the
$\delta$'s in various SUSY breaking scenarions, and investigate
which models pass the phenomenological constraints on $\delta$
parameters. We assume $\delta$'s vanish at some scale (messenger
scale), where soft SUSY breaking terms are generated, and study
the size of the $\delta$'s that are generated by RG evolutions
down to the electroweak scale.
Alternatively, we consider SUSY flavor models where
$\delta$'s are controlled by some flavor symmetry group that acts
on the flavor indices of quarks and their superpartners.

In terms of mass insertion parameters $(\delta_{ij}^d )_{AB}$,
the down-type squark mass matrix of
second and third families can be written as  
\begin{equation}
  \begin{aligned}
  M^2_{\tilde{d}} =
  \left(
    \begin{array}{cccc}
      \widetilde{m}^2_L + \widetilde{m}^2 & 
      \widetilde{m}^2\,\de{23}{LL} & m_s (A_s-\mu \tb) & \widetilde{m}^2\,\de{23}{LR} \\
      \widetilde{m}^2\,\de{23}{LL}^* & 
      \widetilde{m}^2_L + \widetilde{m}^2 &
      \widetilde{m}^2\,\de{23}{RL}^* & m_b(A_b-\mu \tb) \\
      m_s (A_s-\mu \tb) & \widetilde{m}^2\,\de{23}{RL} &
      \widetilde{m}^2_R + \widetilde{m}^2 & \widetilde{m}^2\,\de{23}{RR} \\
      \widetilde{m}^2\,\de{23}{LR}^* & m_b(A_b-\mu \tb) &
      \widetilde{m}^2\,\de{23}{RR}^* & \widetilde{m}^2_R + \widetilde{m}^2
    \end{array}
  \right) ,
  \end{aligned}
\end{equation}
where $\widetilde{m}^2$ is the universal part of soft SUSY
breaking scalar mass squared, and
\begin{equation}
  \begin{aligned}
    \widetilde{m}^2_L = - \frac{1}{6} \cos 2 \beta (m_Z^2+2 m_W^2) , \\
    \widetilde{m}^2_R = - \frac{1}{3} \cos 2 \beta (m_Z^2-m_W^2) ,
  \end{aligned}
\end{equation}
are $D$-term contributions.
We neglect $m_d^2$ terms.
We assume that $A$-terms are
negligible, and the $\mu$ parameter is real.
Relaxing the former assumption is
straightforward, and would not change the results significantly.
The latter assumption is made to satisfy EDM constraints.
By using mass insertion parameters, we have implicitly
specified the basis of squark flavors, i.e.\
the above matrix is in the super CKM basis.
The unitary matrix $U$ diagonalizing the mass matrix
is divided into two parts, $\Gamma_L$ and $\Gamma_R$,
according to the quark chirality to which
they are associated, as
\begin{equation}
  \begin{aligned}
  M^2_{\tilde{d}} &= U^\dagger M^{2\mathrm{(diag)}}_{\tilde{d}}\ U , \\
  \g{L}{I}{j} &\equiv U^I_{\ j} , \\
  \g{R}{I}{j} &\equiv - U^I_{\ j+3} ,
  \end{aligned}
\end{equation}
where $M^{2\mathrm{(diag)}}_{\tilde{d}}$ is a diagonal matrix
with positive elements,
$I = 1,\ldots,6$ is the squark mass eigenstate index, and
$j = 1,2,3$ is the quark mass eigenstate index.
Note that we absorb the relative minus sign between
quark-squark-gluino vertices of opposite chiralities
into that in the definition of $\g{R}{I}{j}$.
We give a name $r_I$
to the ratio of a squark squared mass eigenvalue to the gluino mass squared,
\begin{equation}
  r_I \equiv \frac{\bigl[ M^{2\mathrm{(diag)}}_{\tilde{d}} \bigr]_{II}}
  {m_{\tilde{g}}^2} ,
\end{equation}
which we will use to express Wilson coefficients later on.

\subsection{$\Delta B = 2$ effective Hamiltonian}

For \bsbsbar\ mixing, we use the $\Delta B = 2 (= - \Delta S)$ 
effective Hamiltonian.  We first integrate out SUSY particles and derive 
effective Hamiltonian at sparticle mass scale. Then we use the 
renormalization group running formula from the sparticle mass scale 
to $m_b$ scale presented in 
\cite{Becirevic:2001jj}.
The resulting effective Hamiltonian can be written as 
\begin{equation}
  \mathcal{H}_\mathrm{eff}^{\Delta B = 2} =
  \sum_{i=1}^5 {\cal C}_i Q_i + 
  \sum_{i=1}^3 \widetilde{\cal C}_i \widetilde{Q}_i
  + \mathrm{h.c.},
\end{equation}
where we choose the operator basis as follows: 
\begin{equation}
  \begin{aligned}
    Q_1 & = \bar{s}_L^{\alpha} \gamma_\mu b_L^{\alpha}~
    \bar{s}_L^{\beta}  \gamma^\mu b_L^{\beta} ,
    \\
    Q_2 & = \bar{s}_R^{\alpha} b_L^{\alpha}~\bar{s}_R^{\beta} b_L^{\beta} ,
    \\
    Q_3 & = \bar{s}_R^{\alpha} b_L^{\beta} ~\bar{s}_R^{\beta} b_L^{\alpha} ,
    \\
    Q_4 & = \bar{s}_R^{\alpha} b_L^{\alpha}~\bar{s}_L^{\beta} b_R^{\beta} ,
    \\
    Q_5 & = \bar{s}_R^{\alpha} b_L^{\beta} ~\bar{s}_L^{\beta} b_R^{\alpha} ,
  \end{aligned}
\end{equation}
where $\alpha$ and $\beta$ are color indices.
The Wilson coefficients ${\cal C}_i$'s associated with the operator 
$Q_i$'s are given by 
\begin{equation}
  \begin{aligned}
    {\cal C}_1 = \frac{\alpha_s^2}{216 m_{\tilde{g}}^2}
    &\sum_{IJ} {\g{L}{I}{2}}^* \g{L}{I}{3} {\g{L}{J}{2}}^* \g{L}{J}{3}
    \Bigl( -24 B_2(r_I, r_J) - 264 B_1(r_I, r_J) \Bigr),
    \\
    {\cal C}_2 = \frac{\alpha_s^2}{216 m_{\tilde{g}}^2}
    &\sum_{IJ} {\g{R}{I}{2}}^* \g{L}{I}{3} {\g{R}{J}{2}}^* \g{L}{J}{3}
    \Bigl( -204 B_2(r_I, r_J) \Bigr),
    \\
    {\cal C}_3 = \frac{\alpha_s^2}{216 m_{\tilde{g}}^2}
    &\sum_{IJ} {\g{R}{I}{2}}^* \g{L}{I}{3} {\g{R}{J}{2}}^* \g{L}{J}{3}
    \Bigl( 36 B_2(r_I, r_J) \Bigr),
    \\
    {\cal C}_4 = \frac{\alpha_s^2}{216 m_{\tilde{g}}^2} &\left[
    \sum_{IJ} {\g{R}{I}{2}}^* \g{R}{I}{3} {\g{L}{J}{2}}^* \g{L}{J}{3}
    \Bigl( -504 B_2(r_I, r_J) + 288 B_1(r_I, r_J) \Bigr)
    \right.
    \\ + & \left.
    \sum_{IJ} {\g{L}{I}{2}}^* \g{R}{I}{3} {\g{R}{J}{2}}^* \g{L}{J}{3}
    \Bigl( 528 B_1(r_I, r_J) \Bigr) \right] ,
    \\
    {\cal C}_5 = \frac{\alpha_s^2}{216 m_{\tilde{g}}^2} &\left[
      \sum_{IJ} {\g{R}{I}{2}}^* \g{R}{I}{3} {\g{L}{J}{2}}^* \g{L}{J}{3}
    \Bigl( -24 B_2(r_I, r_J) - 480 B_1(r_I, r_J) \Bigr) \right.
    \\ + & \left.
      \sum_{IJ} {\g{L}{I}{2}}^* \g{R}{I}{3} {\g{R}{J}{2}}^* \g{L}{J}{3}
    \Bigl( 720 B_1(r_I, r_J) \Bigr)
    \right] ,
  \end{aligned}
\end{equation}
where we use the notation
\begin{equation}
  \label{eq:boxloop2}
  B_i (r_I, r_J) = \frac{B_i(r_I) - B_i(r_J)}{r_I - r_J} , \quad
  i = 1,2,
\end{equation}
with \cite{Moroi:2000tk}
\begin{equation}
  \label{eq:boxloop1}
    B_1 (r) = -\frac{r^2\,\ln r}{4(1 - r)^2} - \frac{1}{4(1 - r)} , \quad
    B_2 (r) = -\frac{r\,\ln r}{(1 - r)^2} - \frac{1}{1 - r} .
\end{equation}
One can get $\widetilde{O}_i$ and $\widetilde{\cal C}_i$ for $i =
1,2,3$ by exchanging $L \leftrightarrow R$.

For the matrix elements of the above operators and the numerical values 
of $B_{1,\ldots,5}(\mu)$ and $f_{B_d}$, we use the values given in 
Ref.~\cite{Becirevic:2001jj}. 
We use the following ratio 
\begin{equation}
  \begin{aligned}
    \frac{f_{B_s} \sqrt{B_{B_s}}}{f_{B_d} \sqrt{B_{B_d}}} = 1.21 , 
  \end{aligned}
\end{equation}
given in Ref.~\cite{Okamoto:2005zg}.

\subsection{$\Delta B = 1$ effective Hamiltonian}

Nonleptonic charmless and radiative $B_{d(s)}$ decays are described by
the following $\Delta B=1$ effective Hamiltonian.
We use the same normalization of 
operator basis as in Ref.~\cite{Kane:2002sp}.
RG running of gluino-loop contributions from $m_W$ scale to $m_b$
scale is performed in the way presented in
\cite{Borzumati:1999qt}, i.e., the $\alpha_s^n$ factor from the
quark-squark-gluino vertices is included in an operator rather
than the corresponding Wilson coefficient, and the dimension-five
and dimension-six versions of the (chromo-)magnetic operators are
treated separately. Then the $\Delta B=1$ effective Hamiltonian 
encoding the gluino-squark loop contribution can be written as 
\begin{equation}
  \begin{aligned}
    \mathcal{H}_\mathrm{eff}^{\Delta B = 1} =
    \frac{G_F}{\sqrt{2}}\sum_{p=u,c}
    &\lambda_p \left[
      \sum_{i=3}^6 \left(
      C_i O_i +
      \widetilde{C}_i \widetilde{O}_i
      \right) \right. \\ &+ \left.
      \sum_{i=7\gamma, 8g} \left(
        C_{i b} O_{i b} + C_{i \tilde{g}} O_{i \tilde{g}} +
        \widetilde{C}_{i b} \widetilde{O}_{i b} +
        \widetilde{C}_{i \tilde{g}} \widetilde{O}_{i \tilde{g}} \right)
    \right]
    + \mathrm{h.c.},
  \end{aligned}
\end{equation}
where $\lambda_p = V_{ps}^* V_{pb}$.
The operator basis is chosen as follows:
\begin{equation}
  \begin{aligned}
O_3&= \alpha_s^2\ (\bar s b)_{V-A}
      \sum_q (\bar qq)_{V-A} ,\\
O_4&= \alpha_s^2\ (\bar s_\alpha b_\beta)_{V-A}
      \sum_q (\bar q_\beta q_\alpha)_{V-A} ,\\
O_5&= \alpha_s^2\ (\bar sb)_{V-A}
      \sum_q(\bar qq)_{V+A} ,\\
O_6&= \alpha_s^2\ (\bar s_\alpha b_\beta)_{V-A}
      \sum_q(\bar q_\beta q_\alpha)_{V+A} ,\\
O_{7\gamma b} &= -\frac{\alpha_s\ e}{8\pi^2}\,
                 m_b\,\bar s\,\sigma_{\mu\nu}(1+\gamma_5) F^{\mu\nu} b ,\\
O_{8g b} &= -\frac{\alpha_s\ g_s}{8\pi^2}\,
            m_b\,\bar s\,\sigma_{\mu\nu}(1+\gamma_5) G^{\mu\nu} b ,\\
O_{7\gamma \tilde{g}} &= -\frac{\alpha_s\ e}{8\pi^2}\,
                      \bar s\,\sigma_{\mu\nu}(1+\gamma_5) F^{\mu\nu} b ,\\
O_{8g \tilde{g}} &= -\frac{\alpha_s\ g_s}{8\pi^2}\,
                    \bar s\,\sigma_{\mu\nu}(1+\gamma_5) G^{\mu\nu} b .
  \end{aligned}
\end{equation}
The corresponding Wilson coefficients $C_i$'s are given by
\begin{equation}
  \begin{aligned}
    C_3 = -\frac{1}{2 \sqrt{2} G_F m_{\tilde{g}}^2 \lambda_t} &\left[
    \sum_I {\g{L}{I}{2}}^* \g{L}{I}{3}
    \left( -\frac{1}{18} C_1(r_I) + \frac{1}{2} C_2(r_I) \right) \right. \\ +&
    \left. \sum_{IJ} {\g{L}{I}{2}}^* \g{L}{I}{3} {\g{L}{J}{2}}^* \g{L}{J}{2}
    \left( -\frac{1}{9} B_1(r_I, r_J) - \frac{5}{9} B_2(r_I, r_J) \right)
    \right] , \\
    C_4 = -\frac{1}{2 \sqrt{2} G_F m_{\tilde{g}}^2 \lambda_t} &\left[
    \sum_I {\g{L}{I}{2}}^* \g{L}{I}{3}
    \left( \frac{1}{6} C_1(r_I) - \frac{3}{2} C_2(r_I) \right) \right. \\ +&
    \left. \sum_{IJ} {\g{L}{I}{2}}^* \g{L}{I}{3} {\g{L}{J}{2}}^* \g{L}{J}{2}
    \left( -\frac{7}{3} B_1(r_I, r_J) + \frac{1}{3} B_2(r_I, r_J) \right)
    \right] , \\
    C_5 = -\frac{1}{2 \sqrt{2} G_F m_{\tilde{g}}^2 \lambda_t} &\left[
    \sum_I {\g{L}{I}{2}}^* \g{L}{I}{3}
    \left( -\frac{1}{18} C_1(r_I) + \frac{1}{2} C_2(r_I) \right) \right. \\ +&
    \left. \sum_{IJ} {\g{L}{I}{2}}^* \g{L}{I}{3} {\g{R}{J}{2}}^* \g{R}{J}{2}
    \left( \frac{10}{9} B_1(r_I, r_J) + \frac{1}{18} B_2(r_I, r_J) \right)
    \right] , \\
    C_6 = -\frac{1}{2 \sqrt{2} G_F m_{\tilde{g}}^2 \lambda_t} &\left[
    \sum_I {\g{L}{I}{2}}^* \g{L}{I}{3}
    \left( \frac{1}{6} C_1(r_I) - \frac{3}{2} C_2(r_I) \right) \right. \\ +&
    \left. \sum_{IJ} {\g{L}{I}{2}}^* \g{L}{I}{3} {\g{R}{J}{2}}^* \g{R}{J}{2}
    \left( -\frac{2}{3} B_1(r_I, r_J) + \frac{7}{6} B_2(r_I, r_J) \right)
    \right] , \\
    C_{7\gamma b} = - \frac{\pi}{\sqrt{2} G_F m_{\tilde{g}}^2 \lambda_t}
    &\sum_I {\g{L}{I}{2}}^* \g{L}{I}{3}
    \left( -\frac{4}{9} D_1(r_I) \right)
    , \\
    C_{7\gamma\tilde{g}} = - \frac{\pi}{\sqrt{2} G_F m_{\tilde{g}} \lambda_t}
    &\sum_I {\g{L}{I}{2}}^* \g{R}{I}{3}
    \left( -\frac{4}{9} D_2(r_I) \right)
    , \\
    C_{8\gamma b} = - \frac{\pi}{\sqrt{2} G_F m_{\tilde{g}}^2 \lambda_t}
    &\sum_I {\g{L}{I}{2}}^* \g{L}{I}{3}
    \left( -\frac{1}{6} D_1(r_I) + \frac{3}{2} D_3(r_I) \right)
    , \\
    C_{8\gamma\tilde{g}} = - \frac{\pi}{\sqrt{2} G_F m_{\tilde{g}} \lambda_t}
    &\sum_I {\g{L}{I}{2}}^* \g{R}{I}{3}
    \left( -\frac{1}{6} D_2(r_I) + \frac{3}{2} D_4(r_I) \right)
    .
  \end{aligned}
\end{equation}
One can get $\widetilde{O}_i$ and $\widetilde{C}_i$ for $i =
3,\ldots,6,7\gamma,8g$ by exchanging $L \leftrightarrow R$.
The loop functions are given by
Eqs.~\eqref{eq:boxloop2}, \eqref{eq:boxloop1}, and \cite{Moroi:2000tk}
\begin{equation}
  \begin{aligned}
C_1(r) &= \frac{2 r^3 - 9 r^2 + 18 r - 11 - 6 \ln r}{36 (1 - r)^4} , \\
C_2(r) &= \frac{-16 r^3 + 45 r^2 - 36 r + 7 + 
   6 r^2 (2 r - 3) \ln r}{36 (1 - r)^4} , \\
D_1(r) &= \frac{-r^3 + 6 r^2 - 3 r - 2 - 6 r \ln r}{6 (1 - r)^4} , \\
D_2(r) &= \frac{-r^2 + 1 + 2 r \ln r}{(r - 1)^3} , \\
D_3(r) &= \frac{2 r^3 + 3 r^2 - 6 r + 1 - 6 r^2 \ln r}{6 (1 - r)^4} , \\
D_4(r) &= \frac{-3 r^2 + 4 r - 1 + 2 r^2 \ln r}{(r - 1)^3} .
  \end{aligned}
\end{equation}

\subsection{New elements in this analysis}

SUSY effects in $B_s$ mixing before the CDF/D\O\
measurements of $\Delta M_s$ have been discussed comprehensively
in literatures \cite{Kane:2003zi,Kane:2002sp,Ciuchini:2002uv}.
This work is an update of our previous works \cite{Kane:2003zi,Kane:2002sp}, including 
a few new elements and improvements in the analysis:
\begin{itemize}
\item We include the $\tan\beta$ dependent double mass insertion more 
carefully. As a result, the $B\rightarrow X_s \gamma$ branching ratio 
constrains not only the $LR$ and $RL$ insertions, but also the $LL$ 
and $RR$ insertions, because of the induced LR and RL mass insertions.
Double mass insertion contribution to $\bsg$ has long been known
\cite{Gabbiani:1988rb}.
Potential importance of the double mass insertion was discussed 
in Ref.~\cite{bjkp1} in the context of supersymmetric contributions to
Re $(\epsilon'/\epsilon)$ using the $s\rightarrow d g$ operator,
and similarly in Refs.~\cite{b2s double insertion,Ciuchini:2006dx,Endo:2006dm}
regarding $b \rightarrow s$ transitions.
We discuss more on this in the next subsection in the context
of $b\rightarrow s \gamma$ and $b\rightarrow s g$.
Because of this improvement, we get stronger constraints on the
pure $LL$ or $RR$ insertion, compared with our previous study
\cite{Kane:2002sp}, especially for large $\tan\beta$. 
(However, see also \cite{bsgamma NLO}.)
\item  We also consider the simultaneous presence of the
$LL$ and $RR$ insertions, motivated by some SUSY flavor models
which predict $LL \approx RR$. We find that the $\Delta M_s$
measurement puts a stringent constraint on such cases,
independent of $\tan\beta$ \cite{Gabbiani:1988rb,Ciuchini:2006dx,Endo:2006dm}.
Our analysis shows that some SUSY
flavor models are already excluded by (or marginally compatible
with) the $\Delta M_s$ measurement of the CDF/D\O.
Partly for simplicity, we consider only two cases where
the two insertions are assumed to be correlated by
$\ded{23}{LL} = \pm \ded{23}{RR}$.
Regarding their phases, however, there are good reasons to
restrict their difference around $0$ or $\pi$.
Sizeable $LL$ and $RR$ mass insertions with uncorrelated phases
are likely to give an excessive contribution to the neutron EDM
\cite{neutronEDM}.
For instance,
if $\mu = 500\GeV$ and the sizes of the two insertions are both around 0.05
(see Figs.\ref{fig:LL=RR3}--\ref{fig:LL=-RR10}~(a)),
then the neutron EDM
limits their relative phase within $\lesssim 0.8/\tb$ around 0 or $\pi$.
\item  We include the D\O/CDF data on the phase of $B_s$
mixing deduced from the dilepton charge asymmetry
and $B_s \rightarrow J/\psi \phi$
\cite{Abazov:2008fj,Aaltonen:2007he,phis measurements}.
In particular, we discuss consequences of the present tendency of
the data favoring a negative $O(1)$ value of $\phis$ \cite{Bona:2008jn}.
\item  We present the time dependent CP asymmetry in 
$B^0 \rightarrow K^{*0} \gamma$, in cases
with right-handed $b \leftrightarrow s$ currents
such as from the $RR$ insertion.
See Ref.~\cite{soni} for more details on this observable.
\item  In this paper, we consider only the $LL$ and $RR$ insertions,
and do not consider $LR$ or $RL$ insertion, because the new data
on $\Delta M_s$ does not affect the analysis in Ref.~\cite{Kane:2003zi,Kane:2002sp}
on $LR$ or $RL$ insertion.
In that article, we have found that 
the $\bsg$ constraint
on these chirality-flipping insertions is so strong that
they cannot give an appreciable modification to $\dms$ or $\phis$
\cite{Kane:2002sp,Ciuchini:2002uv}.
\end{itemize}

\subsection{Double mass insertion}

\FIGURE{
  \includegraphics{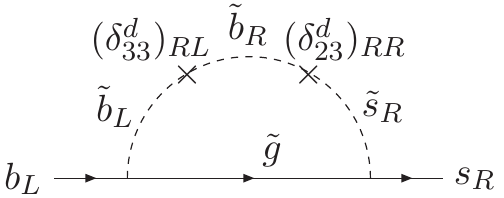}
  \caption{Gluino-squark loop graph with double mass insertion for $\bsg$.}
  \label{fig:bsgamma}
}
If the $LL$ or $RR$ insertion is sizable and $\mu \tan\beta$ is
large, the effective $LR$ or $RL$ insertion can be induced due to
the double mass insertion mechanism we discussed in the previous
subsection and in Refs.~\cite{Gabbiani:1988rb,bjkp1,b2s double insertion}.
Then we can expect that $B\rightarrow X_s \gamma$ could 
give a strong constraint on the $LL$ or $RR$ insertion through
this effective $LR$ or $RL$ insertion. The relevant Feynman
diagram is shown in Fig.~\ref{fig:bsgamma}.
The induced $LR$ or
$RL$ from double mass insertion can be written schematically as
\begin{equation}
( \delta_{LR}^d )_{23}^{\rm ind} =  ( \delta_{LL}^d )_{23} \times
{ m_b ( A_b - \mu \tan\beta ) \over \tilde{m}^2 }.
\end{equation}
Therefore, we have
\[
 ( \delta_{LL,RR}^d )_{23} \sim 10^{-2} \rightarrow
( \delta_{LR,RL}^d )_{23}^{\rm ind} \sim 10^{-2} ,
\]
if $\mu \tan\beta \sim 30\,$TeV\@. This can be expected if $\tan\beta$ is
large $\sim 40$.  For larger $LL,RR$ mixing, even smaller $\mu
\tan\beta$ would  suffice to induce the $LR,RL$ mass insertions of
a size $10^{-2} - 10^{-3}$. Since $\delta_{LL,RR}$'s in SUSY
flavor models are generically  complex, the induced $(
\delta_{LR}^d )_{23}^{\rm ind}$ could carry a new CP violating
phase even if the trilinear coupling $A_b$ and $\mu$ parameters are
real. In such a case, there could be strong correlations among
various CP violating observables. The effects of these induced
$LR$ or $RL$ mixing appear in the deviations in $\SphiK$,
$\ACPbsgam$, or $\SKstargam$ from their SM predictions.  

It is important to remember that the effect of the induced $LR$
insertion is different from that of the single $LR$ insertion,
since they involve different numbers of squark propagators in the
relevant Feynman diagrams, and thus yielding different loop
functions when one evaluates the Feynman diagrams.

\subsection{Numerical analysis}
\label{sec:numerics}

In the following discussions and numerical analysis, we fix the
SUSY parameters as follows once and for all: 
\begin{eqnarray*}
m_{\tilde{q}} = m_{\tilde{g}} = \mu = 500~ {\rm GeV},
\\
\tan\beta = 3 \text{ and } 10, 
\end{eqnarray*}
taking the mass insertion parameters $( \delta^d_{23} )_{AB}$'s as 
a free complex parameter. 
We do not consider very high $\tb \gtrsim 30$ at which
double Higgs penguin contribution may be important
\cite{Foster:2006ze,double higgs penguin}.
Since we do not include the chargino contributions in this work, 
the sign of $\mu$ could be either positive or negative. 
However, we choose a positive $\mu$, since it is preferred by 
the muon $g-2$ when we include the chargino or the neutralino 
contributions.  The plots for a negative $\mu$ are similar to those
for a positive $\mu$.
If the supersymmetric contribution to an observable
is dominated by double mass insertion,
the region allowed by it is almost reflected around zero.
The small difference arises from interference
between single and double insertions.
Such observables include $B(\bsg)$, $\ACPbsgam$, $\SphiK$, and $\SKstargam$.
Therefore,
the compatibility of each case with these observables discussed later
largely remains the same even if we take the negative sign of $\mu$.
When we scan over the complex parameter $( \delta^d_{23} )_{AB}$'s, 
we impose the following constraints and show the excluded regions:
\begin{itemize}
\item Smallest squared mass eigenvalue in $M^2_{\tilde{d}}$
is required to be greater than $(100\ \mathrm{GeV})^2$.
The region incompatible with this requirement is denoted by
gray hatched regions.
\item The branching ratio of $B \rightarrow X_s \gamma$ is required
to be within its $2\sigma$ range \cite{hfag},
  \begin{equation}
    3.0 \times 10^{-4} < B(B \rightarrow X_s \gamma) < 4.1 \times 10^{-4} .
  \end{equation}
The region incompatible with this requirement is denoted by hatched regions.
\item The region allowed by
$12.4\ {\rm ps}^{-1} < \Delta M_s < 23.1\ {\rm ps}^{-1}$ is
denoted by cyan regions.
We allow for up to 30\% of deviation of $\dms$ from the CDF central value
\cite{Abulencia:2006ze},
considering uncertainties in lattice QCD calculation
and the CKM matrix elements
(see e.g.\ \cite{Lenz:2006hd} and references therein).
\item The region allowed by both the $\Delta M_s$ constraint and
$\phis \in [-1.10, -0.36] \cup [-2.77, -2.07]$
\cite{Bona:2008jn},
where $\phis$ is $\arg(M_{12})$, is denoted
by blue regions.
We take the latest 95\% probability range of $\phis$.
We adopt the sign of $\phis$ used in Refs.~\cite{Abazov:2007zj,Dunietz:2000cr}.

\item Then we predict
the time dependent CP asymmetry
($\SphiK$) in $B_d \rightarrow \phi K_S$,
that ($\SKstargam$) in $B^0 \rightarrow K^{*0} \gamma$,
and the direct CP asymmetry ($\ACPbsgam$) in $B \rightarrow X_s \gamma$.
\item A black square denotes the SM prediction for each observable.
\item We show the region corresponding to the $2\sigma$ range of $\SphiK$
in the plots for the allowed
regions in the $({\rm Re} \delta , {\rm Im} \delta )$ plane, using
the current average $\SphiK = 0.39 \pm 0.18$ \cite{hfag}.
For this, we take into account the uncertainty in the prediction of $\SphiK$
coming from the annihilation contribution in QCD factorization \cite{BBNS}
in the same way as in Section~VI.E of Ref.~\cite{Kane:2002sp}.
That is, the prediction of $\SphiK$ from a single point of $\delta$
forms an interval.
We exclude the point of $\delta$ if the interval is mutually exclusive
with the $2\sigma$ range from experiments.
We use this interval in a correlation plot as well.

\end{itemize}

\section{SUSY effects in \boldmath$b\rightarrow s$ after the CDF/D\O\
measurements of $\Delta M_s$}

\subsection{$LL$ insertion case}
\label{sec:LL}

\begin{figure}
  \centering
  \subfigure[\hspace{-5em}]{\includegraphics[height=62mm]{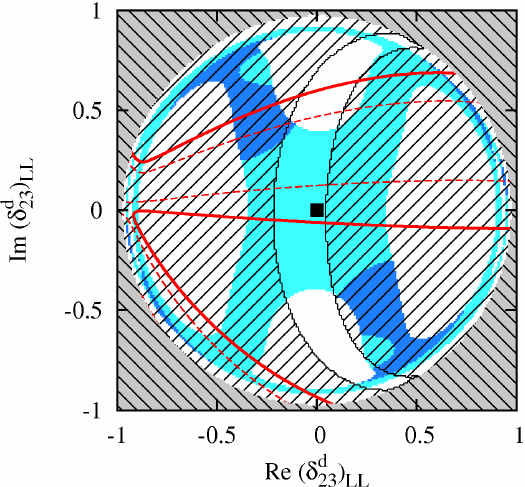}}
  \subfigure[\hspace{-5em}]{\includegraphics[height=62mm]{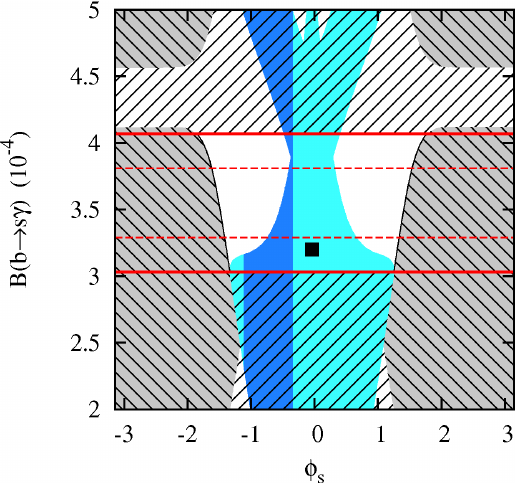}}
  \\
  \subfigure[\hspace{-5em}]{\includegraphics[height=62mm]{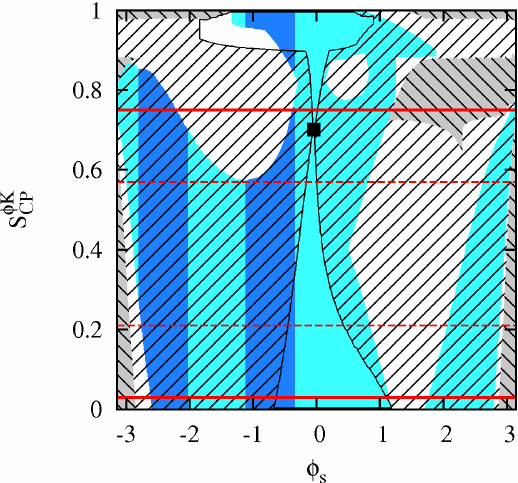}}\hspace{2mm}
  \subfigure[\hspace{-5em}]{\includegraphics[height=62mm]{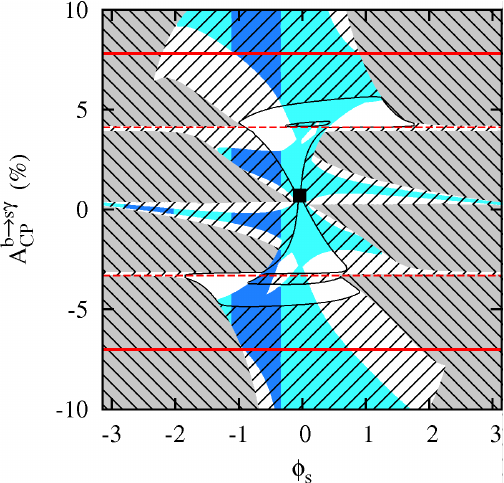}}
\caption{
The $LL$ insertion case with $\tan\beta=3$.
Allowed regions on (a) $({\rm Re} ( \delta_{23}^d )_{LL} ),
{\rm Im} (\delta_{23}^d )_{LL} ) )$, and
correlation between $\phis$ and each of
(b) $B(\bsg)$, (c) $\SphiK$, and (d) $\ACPbsgam$.
The hatched gray region leads to the lightest squark mass $< 100$ GeV.
The hatched region is excluded by the $B\rightarrow X_s \gamma$ constraint.
The cyan region is allowed by $\dms$.
The blue region is allowed both by $\Delta M_s$ and $\phis$.
The black square is the SM point.
In Fig.~(a), bands bounded by red dashed and solid curves
correspond to $1\sigma$ and $2\sigma$ ranges of $S_{\phi K}$, respectively.
In the rest figures, red dashed and solid lines mark
$1\sigma$ and $2\sigma$ ranges of each observable, respectively.}
  \label{fig:LL3}
\end{figure}

Let us first consider the $LL$ insertion (or $LL$ dominance) case 
with $\tan\beta = 3$. 
In the previous study \cite{Kane:2003zi,Kane:2002sp}, we ignored the 
double mass insertion so that the constraint on the $LL$ insertion
was not very strong. In this work, we include the induced $LR$
insertion which is dependent on $\tan\beta$. Therefore, the
$B\rightarrow X_s \gamma$ branching ratio puts a strong
constraint, even before we impose the $\Delta M_s$ measurements.
Only the unhatched region is consistent with $B \rightarrow X_s
\gamma$ constraint in Fig.~\ref{fig:LL3}~(a). A substantial part
of $(\delta^d_{23} )_{LL}$ is already excluded by $B\rightarrow
X_s \gamma$. After imposing the CDF/D\O\ data on $\Delta M_s$ and
$\phis$, only the blue region remains allowed.
It is outstanding that the SM point lies outside the blue region
indicating that the current $\phis$ data, with the aid of $\dms$,
is pointing to a new source of flavor/$CP$ violation.
Moreover, the size of insertion needed to fit the $B_s$ mixing data
is of $O(1)$.
This large insertion inevitably disturbs $\bsg$
through the double mass insertion mechanism involving the $\mu \tan\beta$ term.
Indeed, one finds that most of the blue region is
ruled out by the branching ratio of $\bsg$.
Note that $\bsg$ is this stringent already with $\tb$ as low as 3
and that it grows tighter as $\tb$ increases as we will see shortly.
Still, there are corners compatible with $\bsg$ as well as \bsbsbar\ mixing,
which is evident from Fig.~(b).
The plot also shows that
one of the two $\phis$ solutions is excluded by $\bsg$.
The double insertion leads to
sizable changes in $\SphiK$ or $\ACPbsgam$ as well.
Fig.~(c) shows that $\bsg$ and \bsbsbar\ mixing, together, disfavor
$\SphiK$ around its SM value, although it is
still permitted to fall within its $2\sigma$ range.
The same set of constraints results in $\ACPbsgam$ of $\pm$ a few per cent,
as displayed in Fig.~(d), which can be discriminated from the SM prediction
at a super $B$ factory.

\begin{figure}
  \centering
  \subfigure[\hspace{-5em}]{\includegraphics[height=62mm]{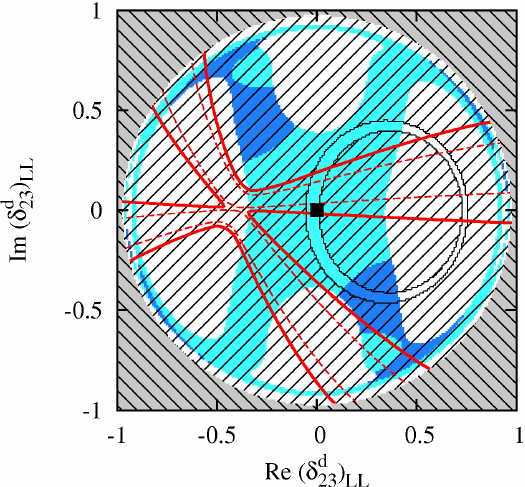}}
  \subfigure[\hspace{-5em}]{\includegraphics[height=62mm]{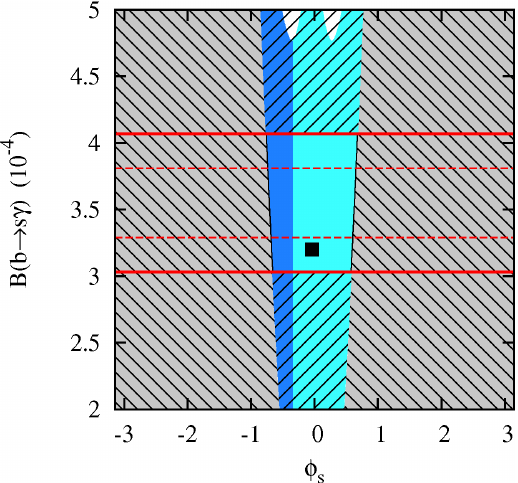}}
  \\
  \subfigure[\hspace{-5em}]{\includegraphics[height=62mm]{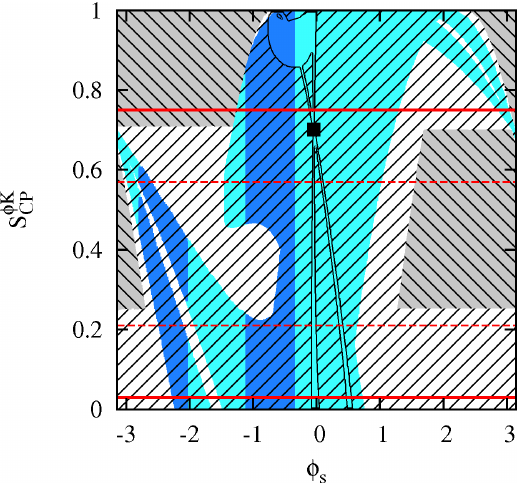}}\hspace{2mm}
  \subfigure[\hspace{-5em}]{\includegraphics[height=62mm]{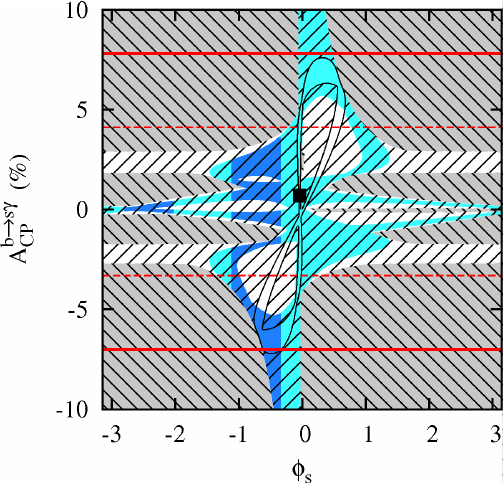}}
  \caption{Plots with the $LL$ insertion for $\tan\beta=10$.
    The meaning of each region is the same as in Figs.~\protect\ref{fig:LL3}.}
  \label{fig:LL10}
\end{figure}

For $\tan\beta = 10$, the double mass insertion becomes more important,
and $( \delta_{23}^d )_{LL}$ is strongly constrained by
$B\rightarrow X_s \gamma$ and $B_s$ mixing constraints. The
results are shown in Figs.~\ref{fig:LL10}.  The
allowed region of $( \delta_{23}^d )_{LL}$ is the narrow unhatched blue strip
in Fig.~(a).
Comparing Figs.~\ref{fig:LL10}~(b) and \ref{fig:LL3}~(b),
one also finds that the phase of \bsbsbar\ mixing is
more tightly constrained compared to the previous case with $\tan\beta=3$.
Also, $\SphiK$ and  $\ACPbsgam$ can deviate from their SM values
significantly through the induced $LR$ insertion.
Fig.~\ref{fig:LL10}~(a) reveals that
the narrow strip allowed by $B_s$ mixing and $\bsg$ leads to
$\SphiK$ out of its $2\sigma$ range.
In this sense, this case with large $LL$ insertion and moderately high $\tb$
is disfavored by the current $B$ physics data.
The predicted range of $\SphiK$ is found
to be higher than its SM value, around 0.9, in Fig.~(c).
In Fig.~(d), we find that the blue unhatched region corresponds to
$\ACPbsgam$ around negative several per cent.

\subsection{$RR$ insertion case}

\begin{figure}
  \centering
  \subfigure[\hspace{-5em}]{\includegraphics[height=62mm]{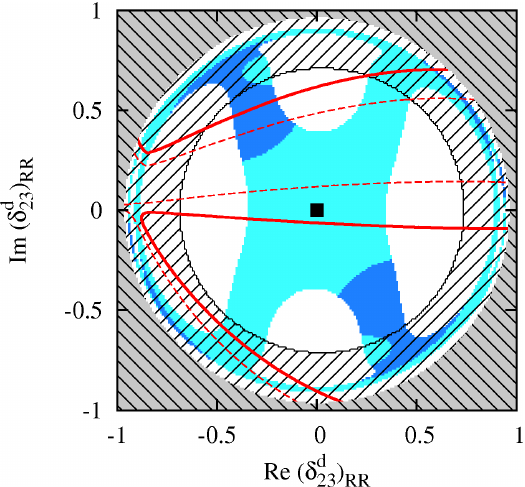}}
  \subfigure[\hspace{-5em}]{\includegraphics[height=62mm]{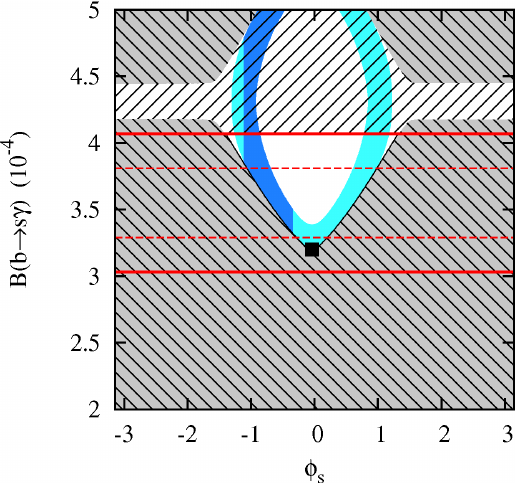}}
  \\
  \subfigure[\hspace{-5em}]{\includegraphics[height=62mm]{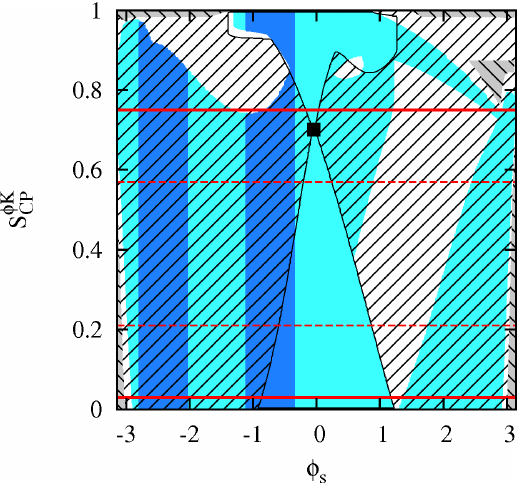}}
  \subfigure[\hspace{-5em}]{\includegraphics[height=62mm]{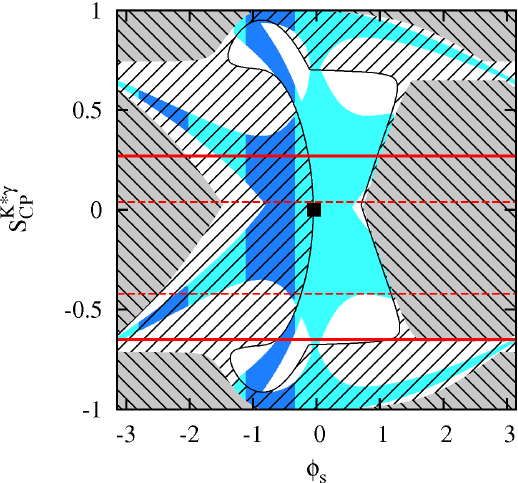}}
\caption{The $RR$ insertion case with $\tan\beta=3$.
    The meaning of each region is the same as in Figs.~\protect\ref{fig:LL3}.}
  \label{fig:RR3}
\end{figure}

\begin{figure}
  \centering
  \subfigure[\hspace{-5em}]{\includegraphics[height=62mm]{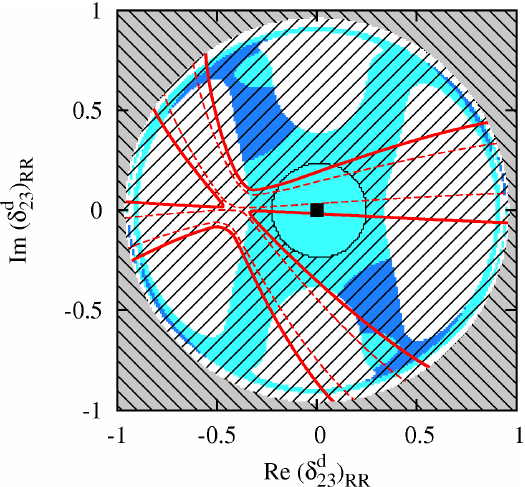}}
  \subfigure[\hspace{-5em}]{\includegraphics[height=62mm]{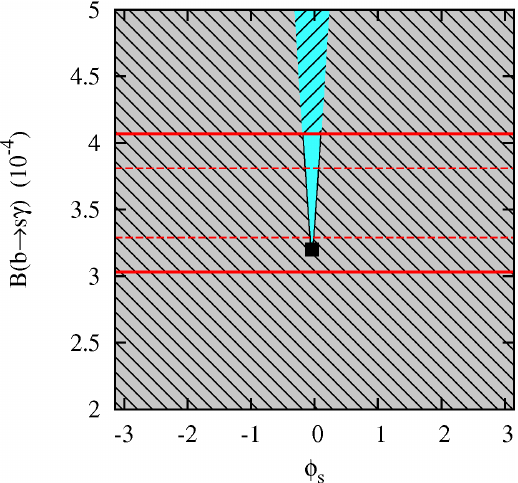}}
  \\
  \subfigure[\hspace{-5em}]{\includegraphics[height=62mm]{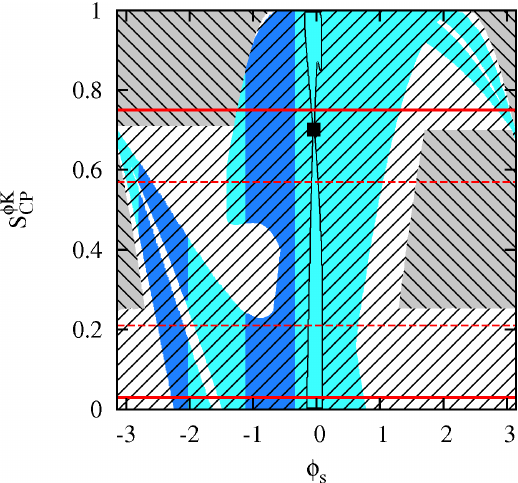}}
  \subfigure[\hspace{-5em}]{\includegraphics[height=62mm]{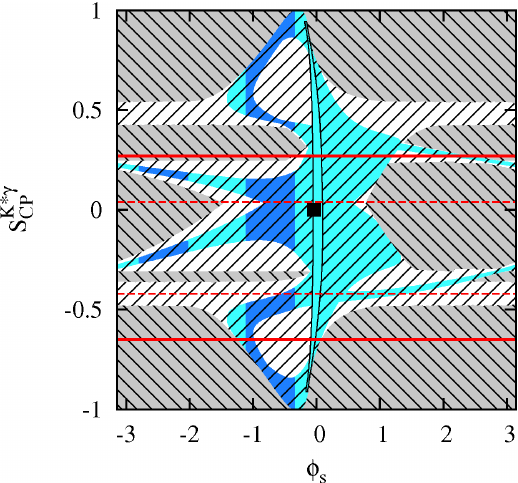}}
  \caption{Plots with the $RR$ insertion for $\tan\beta=10$.
    The meaning of each region is the same as in Figs.~\protect\ref{fig:LL3}.}
  \label{fig:RR10}
\end{figure}

Next, we consider the $RR$ insertion case for $\tan\beta = 3, 10$,
which are shown in Figs.~\ref{fig:RR3}--\ref{fig:RR10}.
The shapes of the allowed regions, after the $B\rightarrow X_s
\gamma$ constraint is imposed, are different from those in the $LL$
insertion case, since there is no interference between the SUSY
amplitude (the original $RR$ or the induced $RL$ type) and the SM
amplitude ($LR$ type). However, the general tendency is similar to
the LL insertion case: namely, the induced $RL$ insertion
involving the double mass insertion is constrained by the
$B\rightarrow X_s \gamma$ branching ratio, and the constraint
becomes severer for larger $\tan\beta$.

In Fig.~\ref{fig:RR3}, the $\dms$ and $\phis$ constraints again
excludes the origin and requires nonzero squark mixing depicted by
the blue region.
We observe that $\bsg$ leaves a broader region than in the $LL$ case
(compare Figs.~\ref{fig:RR3}~(a) and \ref{fig:LL3}~(a)).
In particular, there remains a larger portion of unhatched blue region,
due to the weaker constraint from $\bsg$.
Still, only one of the two solutions of $\phis$
is allowed in Fig.~\ref{fig:RR3}~(b).
The induced $RL$ insertion can lead to
sizable changes in $\SphiK$ and/or $\SKstargam$ as well,
as shown in Figs.~(c) and (d).
Each of them deviates from its SM value due to
the $O(1)$ phase of $\ded{23}{RR}$ favored by $\phis$,
under the $\bsg$ constraint.
Although $\SphiK$ is expelled from the SM point, it can still remain
consistent with its measurements.
Note that $\SKstargam$ could be as large as around $\pm 0.8$, and
these values are in fact preferred by $\phis$ and $\bsg$.
This would be clearly tested at $B$ factories.

For $\tan\beta = 10$, the double mass insertion becomes more important,
and $( \delta_{23}^d )_{RR}$ is strongly constrained by both
$B\rightarrow X_s \gamma$ and $B_s$ mixing. The
results are shown in Fig.~\ref{fig:RR10}~(a). In this case, the
region of $( \delta_{23}^d )_{RR}$ allowed by $\bsg$ and $\dms$
is smaller than the previous case with $\tan\beta=3$.
Moreover, the limitation is so strong that the measured value of $\phis$
cannot be reached.
Therefore, this case with large $RR$ insertion and moderately high $\tb$
is disfavored by the current $B$ physics data.
Indeed, $\phis$ is confined within a narrow range around the SM value
and thus no unhatched blue region can be found in Fig.~(b).
Forgetting about the current status of $\phis$,
one might estimate effects of the $RR$ insertion
within the unhatched cyan region on $\SphiK$ and $\ACPbsgam$.
They may deviate from their SM values significantly through
induced $RL$ insertion, as shown in Figs.~(c) and (d).

\subsection{$LL = RR$ case}

\begin{figure}
  \centering
  \subfigure[\hspace{-5em}]{\includegraphics[height=62mm]{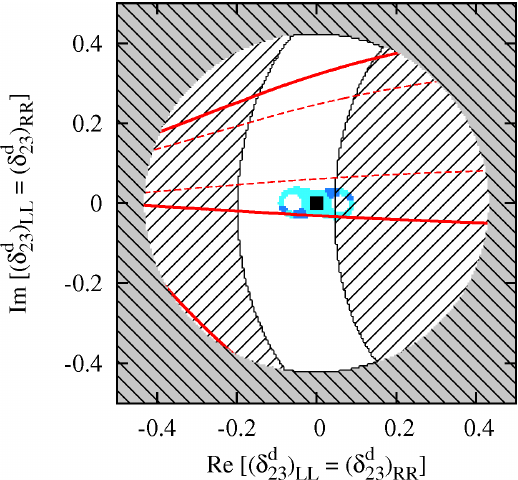}}
  \subfigure[\hspace{-5em}]{\includegraphics[height=62mm]{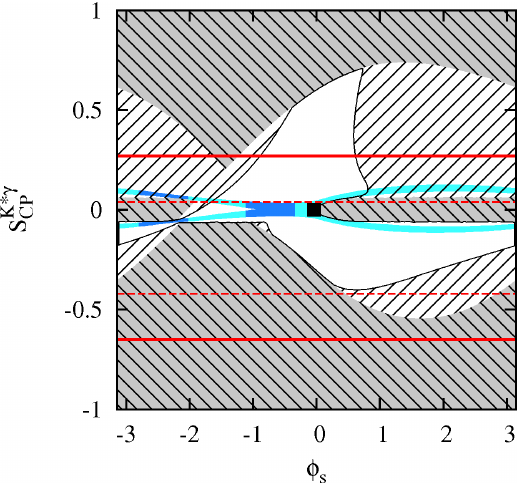}}
  \\
  \subfigure[\hspace{-5em}]{\includegraphics[height=62mm]{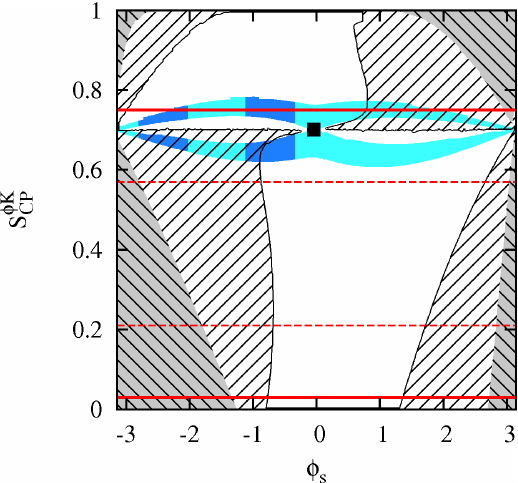}}\hspace{2mm}
  \subfigure[\hspace{-5em}]{\includegraphics[height=62mm]{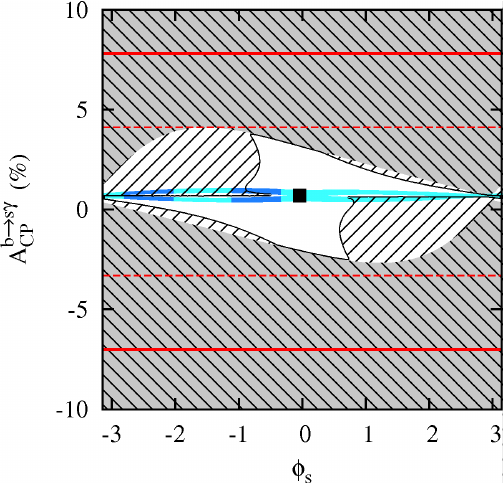}}
  \caption{Plots for the $LL=RR$ case with $\tan\beta=3$.
    The meaning of each region is the same as in Figs.~\protect\ref{fig:LL3}.}
  \label{fig:LL=RR3}
\end{figure}

\begin{figure}
\centering
\subfigure[\hspace{-5em}]{\includegraphics[height=62mm]{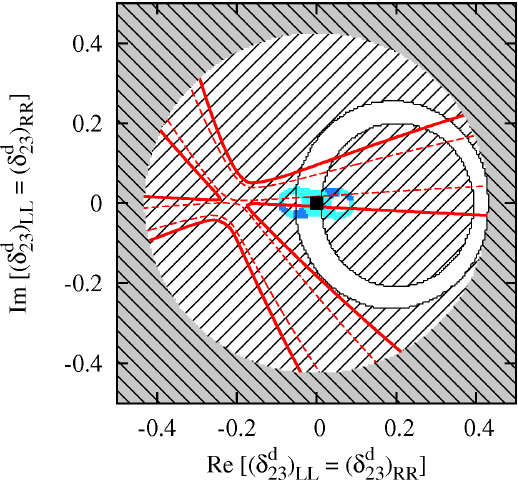}}
\subfigure[\hspace{-5em}]{\includegraphics[height=62mm]{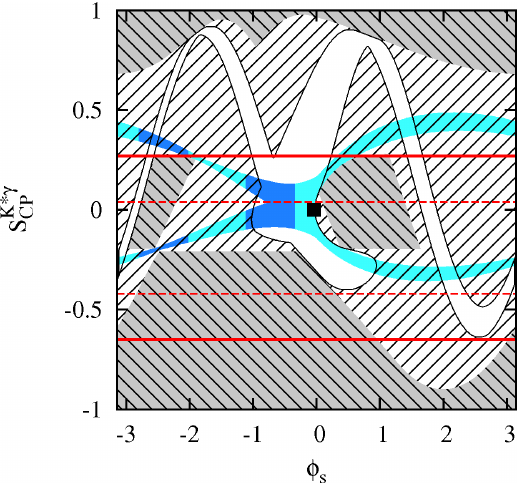}}
\\
\subfigure[\hspace{-5em}]{\includegraphics[height=62mm]{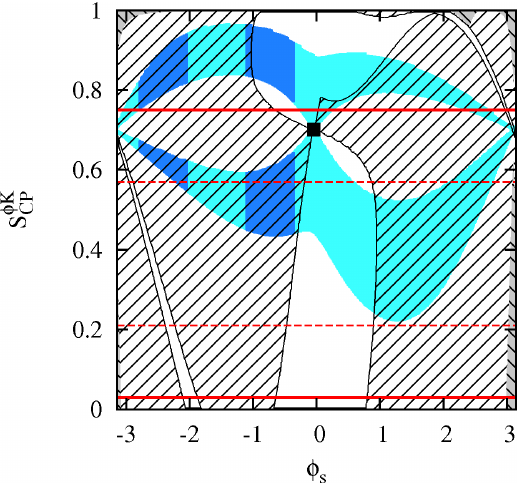}}\hspace{2mm}
\subfigure[\hspace{-5em}]{\includegraphics[height=62mm]{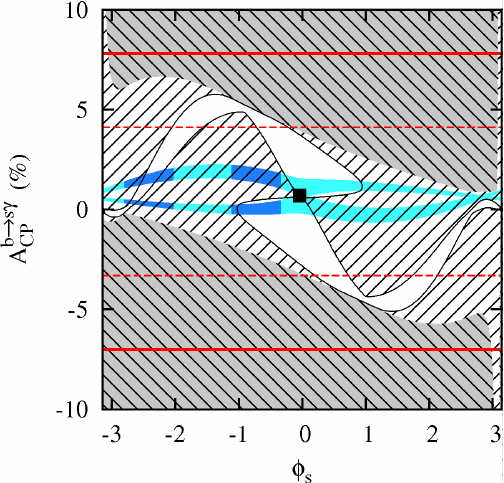}}
\caption{Plots for the  $LL=RR$ case with $\tan \beta = 10$.
    The meaning of each region is the same as in Figs.~\protect\ref{fig:LL3}.}
\label{fig:LL=RR10}
\end{figure}

In this section, we consider the $LL=RR$ case with $\tan\beta = 3,
10$, which are shown in Figs.~\ref{fig:LL=RR3} and \ref{fig:LL=RR10},
respectively.  In this case,
the supersymmetric effect on \bsbsbar\ mixing is greatly enhanced 
compared to the $LL$ or the $RR$ insertion case,
while that on $\bsg$ is not.
Thus, only a tiny region around zero is allowed even for
small $\tan\beta =3$, shown in Fig.~\ref{fig:LL=RR3}~(a).
The phase of the mixing is not
constrained significantly by $\bsg$, and this decay alone allows
for an arbitrary $\phis$,
as can be seen in the other three plots.
These plots also show variations in $\SKstargam$, $\SphiK$, and $\ACPbsgam$,
but they are much smaller than
are found in the preceding cases with a single insertion of either chirality,
since $\dms$ allows a much smaller squark mixing.
This means that this case can account for the current data of
$\phis$ as well as $\dms$ while obeying the other constraints
on $CP$ asymmetries under consideration.
Still, differences of $\SKstargam$ and $\SphiK$
from their SM predictions can be comparable to or larger than
their sensitivities at a super $B$ factory,
while $\ACPbsgam$ is not altered enough.
Note that the blue region again implies a non-vanishing
discrepancy in $\SphiK$.

The results for a higher $\tan\beta =10$ are shown in Fig.~\ref{fig:LL=RR10}.
The $\bsg$ constraint becomes stronger.
Because of this, the range of $\phis$ is reduced, but
it can still be consistent with the present data.
Also, the increased effect of the double insertion
leads to larger deviations in $\SKstargam$, $\SphiK$, and $\ACPbsgam$.
In particular, one finds that the unhatched blue region
leading to $\SphiK \sim 0.9$ is excluded
from its $2\sigma$ band in Fig.~(c).
Therefore this case is disfavored by the current data.
Note that the SUSY effect in $\SphiK$ depends on the sum of the
$LL$ and $RR$ (or $LR$ and $RL$) insertions, and this makes a clear difference
between the predictions of $CP$ asymmetry in this case and the next.

\subsection{$LL = -RR$ case}

\begin{figure}
  \centering
  \subfigure[\hspace{-5em}]{\includegraphics[height=62mm]{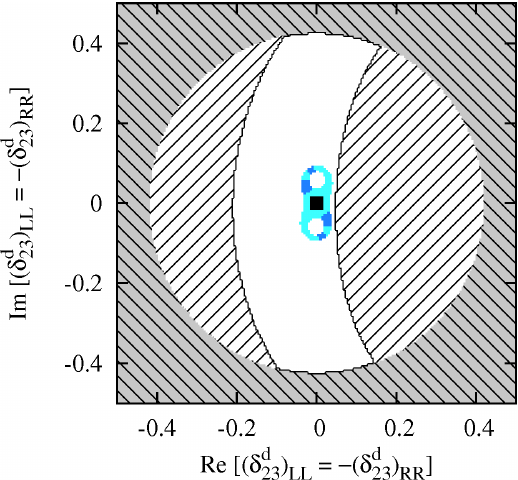}}
  \subfigure[\hspace{-5em}]{\includegraphics[height=62mm]{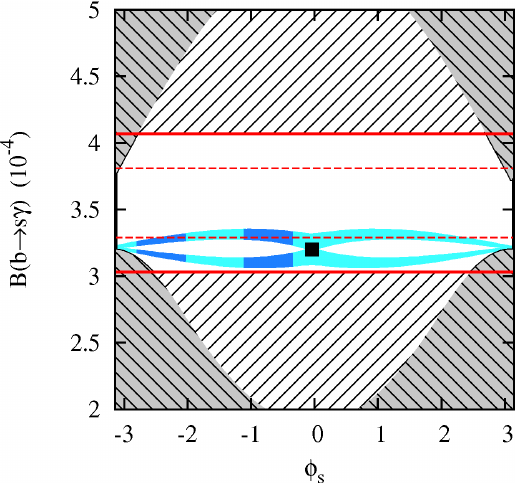}}
  \\
  \subfigure[\hspace{-5em}]{\includegraphics[height=62mm]{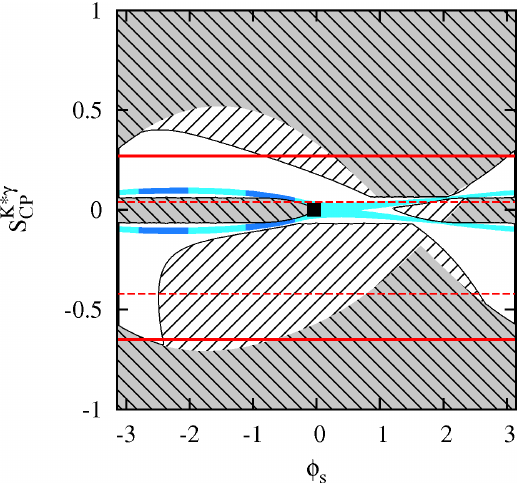}}\hspace{2mm}
  \subfigure[\hspace{-5em}]{\includegraphics[height=62mm]{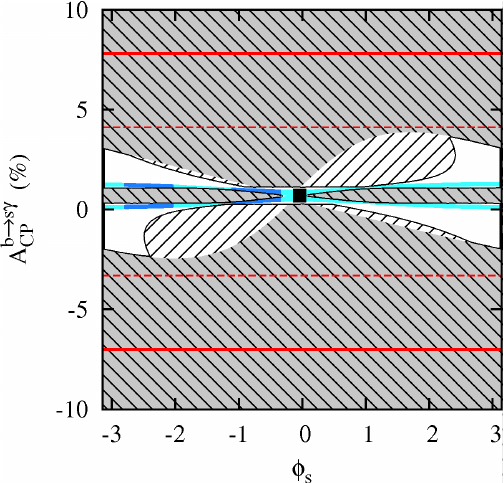}}
  \caption{Plots for the $LL=-RR$ case with $\tan\beta=3$.
    The meaning of each region is the same as in Figs.~\protect\ref{fig:LL3}.}
  \label{fig:LL=-RR3}
\end{figure}

In this section, we consider the $LL=-RR$ case.
The results for $\tb = 3$ are shown in Figs.~\ref{fig:LL=-RR3}.
Note that the \bsbsbar\ mixing constraint is again much
stronger than 
a case with a single insertion of either chirality,
and only a tiny region around zero is allowed.
The phase of the mixing can be arbitrary even after $\bsg$ has been imposed,
as is shown in Fig.~(b).
The deviation in $\SKstargam$ can be comparable to or larger than
its sensitivities at a super $B$ factory,
while $\ACPbsgam$ is not altered enough.
In this case, $\SphiK$ does not move from its SM value,
as the SUSY effect in $\SphiK$ depends on the sum of the
$LL$ and $RR$ (or $LR$ and $RL$) insertions which cancel each other.
Therefore $\SphiK$ is not affected even for higher $\tb$.
Instead, $\SetapK$ should show a discrepancy as it depends on
the difference of the $LL$ and $RR$ (or $LR$ and $RL$) insertions.

\begin{figure}
 \centering
  \subfigure[\hspace{-5em}]{\includegraphics[height=62mm]{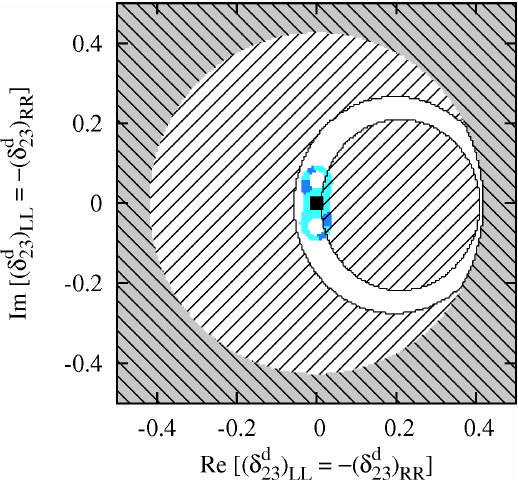}}
  \subfigure[\hspace{-5em}]{\includegraphics[height=62mm]{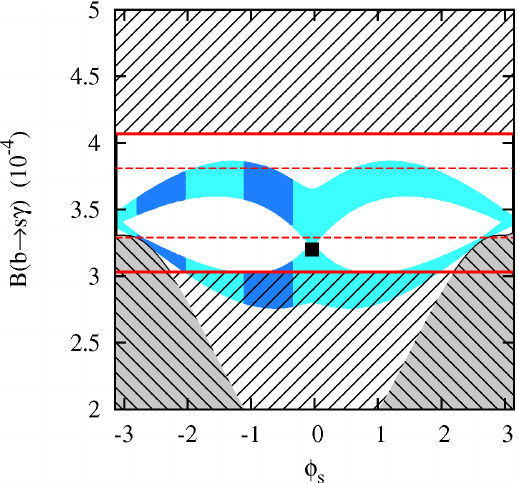}}
  \\
  \subfigure[\hspace{-5em}]{\includegraphics[height=62mm]{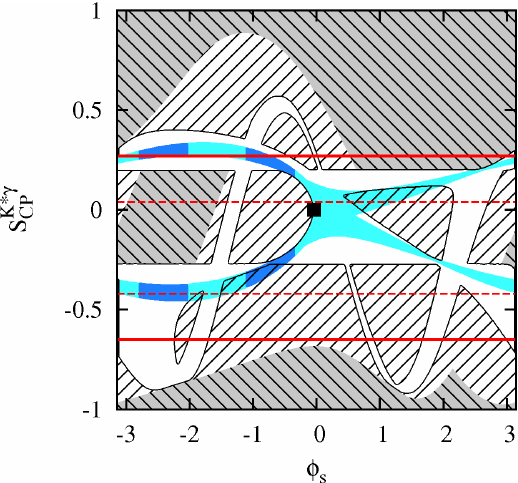}}\hspace{2mm}
  \subfigure[\hspace{-5em}]{\includegraphics[height=62mm]{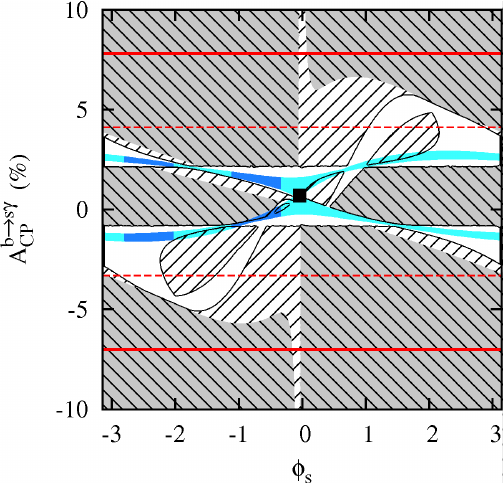}}
  \caption{Plots for the  $LL=-RR$ case with $\tan \beta = 10$.
    The meaning of each region is the same as in Figs.~\protect\ref{fig:LL3}}
\label{fig:LL=-RR10}
\end{figure}

Results for a higher $\tan\beta =10$ are shown in Figs.~\ref{fig:LL=-RR10}.
The $\bsg$ constraint becomes stronger.
Nevertheless, $\phis$ is allowed to have an arbitrary value.
Deviations in $\SKstargam$ and $\ACPbsgam$ has been amplified
relative to the previous case with $\tb=5$.
As was mentioned above, $\SphiK$ remains at its SM prediction.
This helps the present case to be compatible
with all of the experimental inputs,
$\bsg$, $\dms$, $\phis$, and $\SphiK$,
even for a moderately high $\tb$.
Recall that the $LL = RR$ case, by contrast, was in conflict with $\SphiK$
for the same value of $\tb$.
The phase of mass insertions in the unhatched blue region
causes non-vanishing deviations in $\SKstargam$ and $\ACPbsgam$,
to such an extent that can be tested at a $B$ factory.

\subsection{Implications for $B_s \rightarrow \mu^+ \mu^-$}

In the previous sections, we derived the constraints on the $LL$
and $RR$ insertions related to the $23$ mixing in the squark sector.
Gluino mediated flavor violation of $b \rightarrow s$ can affect
another rare $B_s$ decay, $B_s \rightarrow \mu^+ \mu^-$. Isidori
and Retico obtained bounds on $\delta_{LL,RR}$'s from
$B( B_s \rightarrow \mu^+ \mu^- )$ for light $m_A \approx 200$ GeV
and large $\tan\beta$ \cite{isidori}:
\begin{equation}
    {{{B ( B_s \rightarrow \mu^+ \mu^- ) } \over {B_{\rm SM}
    (B_s \rightarrow \mu^+ \mu^- )}} \approx 1.5 \times
    10^5~\left( { 200 \over M_A (GeV) } \right)^4~| (
    \delta_{23}^d )_{LL,RR} |^2~{ \left( {\tan\beta \over 50 }
    \right)^6 \over { \left[ {2\over 3} + {1 \over 3}~
    \left( \tan\beta \over 50 \right) \right]^4 }  }}
\end{equation}
Our constraint is independent of $m_A$, and is mainly driven by $B
\rightarrow X_s \gamma$ for large $\tan\beta$ case, where $B_s
\rightarrow \mu^+ \mu^-$ can be enhanced.

Note that $| \delta_{LL,RR} | \lesssim 0.05 $ for $\tan\beta=40$ from 
\bsbsbar\ mixing.  
This constraint and the ones considered in the preceding sections
are complementary with each other. For small and
moderate $\tan\beta \lesssim 30$, the constraint derived in this
work from $B\rightarrow X_s \gamma$ and $\Delta M_s$ is more
important than that from $B_s \rightarrow \mu^+ \mu^-$.

The gluino contributions to $B_s \rightarrow \mu^+ \mu^-$ is not
that important in general, unless $m_A$ is light and
$\tan\beta$ is very large.

\section{Implications for SUSY models}

In Sec.~3, we derived the constraint on $\ded{23}{LL}$ and $\ded{23}{RR}$.
The size of
$\delta$'s are determined by theories for the soft SUSY breaking,
or SUSY breaking mediation mechanisms. There are basically three
categories in the solutions to the SUSY flavor and CP problems:
\begin{itemize}
\item Universal scalar masses at some messenger scale

\item Alignment of quark and squark mass matrices in the flavor
space using some flavor symmetry

\item Decoupling (effective SUSY scenario).
\end{itemize}
In this section, we discuss implications of
the analysis in the previous section 
on the flavor structures of the soft terms at high energy scale and on SUSY
flavor models, for the first two categories listed above
to which our results are applicable.

\subsection{SUSY models with universal scalar masses}

Let us first discuss the flavor physics within SUSY scenarios
where one has universal soft terms at some high energy messenger scale
$M_{\rm mess}$. In this case, the SUSY flavor problem is solved by
assuming universal squark mass matrices at $M_{\rm mess}$.
Nonetheless at electroweak scale, non-vanishing mass
insertion parameters are generted by RG evolution, which is
calculable in terms of the Yukawa couplings. Namely, $\delta_{ij}
( M_{\rm mess} ) = 0$, and nonzero $\delta$'s at the
electroweak scale are generated by RG evolutions. Models belonging to
this category include the so-called minimal supergravity (mSUGRA)
or gauge mediation SUSY breaking scenarios, dilaton dominated SUSY
breaking within superstring models.

For example, within mSUGRA, one has \cite{sugra_ll} 
\begin{equation}
( \Delta_{ij} )_{LL} ( M_Z ) \simeq -{1 \over 8
\pi^2}\,Y_t^2 \left( V_{\rm CKM} \right)_{3i} \left( V_{\rm CKM}^*
\right)_{3j}\, \left( 3 m_0^2 + a_0^2 \right)\, \log ( { M_{*}
\over M_Z } ) , \label{eq:msugraLL}
\end{equation}
so that $( \delta_{LL}^d )_{23} \simeq 10^{-2}$ and $(
\delta_{LL} )_{13} \simeq 8 \times 10^{-3} \times e^{- i 2.7}$.
This size of $(\delta^d_{23} )_{LL}$ might be regarded as being
perfectly fine with the constraints we discussed in Subsection~\ref{sec:LL},
unless one cares about the current status of $\phis$.
If one is interested in fitting the present data of $\phis$,
this scenario is not a good choice.
In particular, the phase of $\ded{23}{LL}$ is $-0.02$.
Therefore, there would be only small deviations in
$\phis$, $\SphiK$, or $\ACPbsgam$ within this scenario.
There could be some effects in $b\rightarrow d$ transition, including
$B\rightarrow X_d \gamma$, and we refer to Ref.~\cite{kramer} for
further details.

If we consider a SUSY grand unified theory (GUT) with right handed neutrinos,
the situation can change, however. In many SUSY GUT models,
the left handed lepton doublet sits in the same representation as
the left-handed anti-down quark triplets.
Then, the large mixing in the atmospheric neutrinos could be related with
the large mixing in the $\widetilde{b}_R$--$\widetilde{s}_R$ sector
\cite{Moroi:2000tk,Moroi:2000mr,Chang:2002mq},
unless the main source of neutrino mixings is
the Majorana right-handed neutrino mass terms.
Therefore there could be large $b\rightarrow s$ transitions in the low energy
processes  in such scenarios, and $B_s$ mixing or
$B_d \rightarrow \phi K_S$ CP asymmetries can differ significantly
from the SM predictions.

For example, in SU(5) with right-handed neutrinos,
one has \cite{Moroi:2000tk,Moroi:2000mr}
\[
  \begin{aligned}
    ( 
  m^2_{\tilde{d}} )_{ij} &\simeq - \frac{1}{8 \pi^2}
  [Y_N^\dagger Y_N ]_{ij}  ( 3 m_0^2 + A^2 ) \log \frac{M_*}{M_{\rm GUT}}
  \\
  &
  \simeq - e^{-i (\phi^{(L)}_i - \phi^{(L)}_j)}
  {y_{\nu_k}^2 \over 8\pi^2}~[V_L^*]_{ki} [V_L]_{kj} ( 3 m_0^2 + A^2 )
  \log \frac{M_*}{M_{\rm GUT}} .
  \end{aligned}
\]
In this scenario, $| ( \delta^d_{RR} )_{23} | \simeq 2 \times
10^{-2} \times \left( {M_{N_3} / 10^{14}~{\rm GeV}} \right) $ with
$O(1)$ phase, which is in sharp contrast with the $LL$ insertion,
Eq.~(\ref{eq:msugraLL}). 
This RG induced $\delta$ alone is
small enough to evade the constraint from $\dms$, but
not big enough to accommodate $\phis$.
On the other hand, the $RR$ insertion is large enough to
induce an effective $RL$ insertion of $\sim 10^{-2}$ through the
double mass insertion mechanism, and can affect
$\SphiK$ and $\SKstargam$.
Also in this scenario,
there are RG induced $LL$ insertions mentioned above.
Combining these two types of insertions,
one could get enough effect in \bsbsbar\ mixing
to fit the current world average of $\phis$.
However, an obstacle to this purpose is hadronic electric dipole moment
\cite{fcncgut}.
In particular, it is not easy to circumvent this constraint
if one assumes that the $LL$ insertion arises solely from RG evolution,
as is the case in this subsection.
One of the few ways might be to assume that the first and the second terms
in $A_b - \mu \tb$ cancel each other resulting in a small sum,
since the supersymmetric contribution to hadronic electric dipole moment
is proportional to the sum.

\subsection{SUSY flavor models}

Another way out of the SUSY flavor problem is to invoke some flavor symmetry
and make quark and squark mass matrices almost aligned.
Alignment of quark and squark mass matrices can be achieved by assuming
some flavour symmetries ( $U(1), S_3,$.... ).
We discuss what implications the present analysis may have on
those supersymmetric flavor models.
We borrow the list of models from Ref.~\cite{Randall:1998te},
discarding two decoupling type models therein.

Suppose that a given flavor symmetry is broken around the GUT scale.
Then RG evolution of the squark mass matrix down to the weak scale
should be taken into account.
The diagonal components increase receiving the gluino mass contribution:
\begin{equation}
  \label{eq:diagonalrunning}
  m^2_{\widetilde{q}}\, (M_Z) \approx m^2_0 + 6\, M_{1/2}^2 ,
\end{equation}
where $m_0$ and $M_{1/2}$ are the diagonal squark mass and the gluino mass
at the GUT scale.
An off-diagonal element does not change very much except for
the CKM suppressed contribution in Eq.~\eqref{eq:msugraLL}.
In many cases, a flavor symmetry predicts the ratio of
an off-diagonal element to the diagonal one,
$(\Delta_{ij})_{AB} / m^2_0$, thereby determining the degree of
squark non-universality at the scale where it is broken.
In terms of this ratio, the mass insertion at weak scale can be written as
\begin{equation}
\ded{ij}{AB} \approx \frac{(\Delta_{ij})_{AB}/m^2_0}{1 + 6\,M_{1/2}^2 / m^2_0},
\end{equation}
using Eq.~\eqref{eq:diagonalrunning}.
One can notice that
the non-universality at the GUT scale is diluted in the course of running,
depending on the ratio $M_{1/2}^2 / m^2_0$.
In what follows, we ignore this effect.
If one takes it into account, constraints on a model may be eased
especially for large $M_{1/2}$.
On the other hand, this could also make it more difficult
to account for the present $O(1)$ value of $\phis$
by reducing the expected size of a mass insertion below what is needed.

The result is shown in Table~\ref{tab:models}.
\TABLE{
  \renewcommand{\arraystretch}{1.1}
  \begin{tabular}{ccccc}
    \hline
    Model & 
    $|\ded{23}{LL}|$ & 
    $|\ded{23}{RR}|$ & 
    $\tb = 3$ &
    $\tb = 10$
    \\
    \cite{RS32} &
    $\lambda^2$ & $\lambda^4$ &
    $\cdot$ &
    $\surd$
    \\
    \cite{RS33}, \cite{RS35}a &
    $\lambda^2$ & $1$ &
    $\times$ &
    $\times$
    \\
    \cite{RS34} &
    $\lambda^2$ & $\lambda^8$ &
    $\cdot$ &
    $\surd$
    \\
    \cite{RS35}b &
    $\lambda^2$ & $\lambda^{1/2}$ &
    $\times$ &
    $\times$
    \\
    \cite{RS37}, \cite{RS40}b &
    $\lambda^2$ & $\lambda^2$     &
    $\phis$&
    $\surd$
    \\
    \cite{RS38} &
    $\lambda^3$ & $\lambda^5$     &
    $\cdot$ &
    $\cdot$
    \\
    \cite{RS42} &
    $\lambda^2$ & $\lambda^4$     &
    $\cdot$ &
    $\surd$
    \\
    \hline
  \end{tabular}
  \caption{Status of
    part of the models analyzed in Ref.~\cite{Randall:1998te},
    for the two different values of $\tb$.
    Each case is classified into one of the following four categories:
    ($\cdot$) incompatible with $\phis$ but safe otherwise;
    ($\phis$) compatible with $\phis$ and safe;
    ($\surd$) currently okay but dangerous;
    ($\times$) disfavored.
  }
  \label{tab:models}
}%
The current status of each model is indicated in the two columns on the right.
One can see that availability of the new data on \bsbsbar\ mixing
enables us to discriminate models
according to their predictions on 2--3 mixing of down-type squarks.
A Model is marked as being safe
if it suppresses flavor violation to such an extent that
no appreciable deviation from the SM can be observed.
However, such a model may not produce enough difference in $\phis$
to account for its current world average.
We indicate a class of models that can fit $\phis$
while keeping compatibility with the other constraints.
They lead to nonzero mass insertions of both chiralities
enhancing supersymmetric contribution to $B_s$ mixing.
A caveat is dilution of mass insertions mentioned above.
Some models leading to sizeable mass insertions
are about to be in contact with the present experiments
or strongly disfavored by them depending on the choice of parameters.
A future experiment should be able to resolve this issue
and to scrutinize more models.
Needless to say, all the above discussions are based on our choice
of sparticle mass scale.
That is, supersymmetric flavor/$CP$ problems can be mitigated
by making sparticles heavier.

\section{Conclusions}

In conclusions, we studied the implication of the recent
measurements of $B_s - \overline{B_s}$ mixing on the mass
insertion parameters in the general SUSY models and on the SUSY
flavor models. The recent measurements of $\dms$
constrains the CKM element $| V_{td} |$, which is
consistent with the Belle result extracted from $b\rightarrow d
\gamma$. This constitutes another test of the CKM paradigm of the
SM for flavor and CP violation in the quark sector. The
measurement of $\Delta M_s$ begins to put strong constraint on new
physics scenarios, and a room for the new physics contribution to $b
\rightarrow s$ transition is getting tight now, and will be even
more so in the future. Even the very first data on $\Delta M_s$
from D\O\ and CDF already constrain either of the $LL$ and
the $RR$ insertions, 
which should be compared
with the bounds $\lesssim O(1)$ in \cite{Kane:2002sp} or \cite{Ciuchini:2002uv}. For
the $LL = \pm RR$ case, the constraints are even stronger,
and the allowed mass insertion parameters are tiny even for small
$\tan\beta =3$. Still there could be moderate to large deviations
in $A_{\rm CP}^{b \rightarrow s \gamma}$, $\SKstargam$, or $\SphiK$
through the double mass insertion
effects for large $\tan\beta$ case. It is imperative to measure
these observables, and confirm the SM predictions on these
observables both at hadron colliders and at (super) $B$ factories,
in order to test the CKM paradigm in the $b\rightarrow s$
transition. 

In a model independent approach, one can say that
CP violation in $B_s \rightarrow J/\psi \phi$ and $A_{\rm SL}$  give
additional informations on the phase of \bsbsbar\ mixing,
and can make a firm test of the CKM paradigm in the SM, and constrain
various new physics scenarios.
CP asymmetries in $B\rightarrow \phi K_S , \eta' K_S, K_S \pi^0$,...
can differ from the SM predictions to some extent, but we cannot make
definite predictions within the model independent appraoch.

Within general SUSY models with gluino mediated $b\rightarrow s$ transition,
one can summarize the implications of the $\Delta M_s$ and $\phis$ measurements
as follows:
\begin{itemize}
\item The $LL$ or $RR$ insertions for small $\tan\beta$ case
cannot be large as in the past ($\lesssim 0.5$)
\item Large $\tan\beta$ case is strongly contrained by
$b\rightarrow s\gamma$ (independent of $m_A$) and by
$B_s \rightarrow \mu^+ \mu^-$ for light $m_A$ 
\item The $LL= \pm RR$ case is even more strongly contrained 
by $\Delta M_s$ measurement
\item The $LR$ or $RL$ insertions consistent with $b\rightarrow s \gamma$
is still fine with $\dms$,
since it does not affect \bsbsbar\ mixing;
however for the same reason, it cannot make an $O(1)$ difference in $\phis$
\item Definite relations between $\Delta B=2$ and  $\Delta B=1$ processes
CP asymmetries in $B\rightarrow \phi K_S , \eta' K_S, K_S \pi^0$,...
can differ from the SM predictions to some extent, and we can make
definite predictions within SUSY models (modulo hadronic uncertainties)
\item
$B_d \rightarrow \phi K_S $ can still differ from the SM prediction,
if the (induced) $LR$ or $RL$ insertions are present at the level of
$10^{-2}$--$10^{-3}$
\end{itemize}
Whether the present hint of new physics in $B_s$ mixing phase
will persist in the future or not
will be an intersting topic within coming years for B factories and hadron
colliders, and the data will show whether the SM explains
$b\rightarrow s$ transition perfectly, or some new physics is in need.
In particular it is important to improve the precision of
time dependent CP asymmetries in $B_s \rightarrow J/\psi \phi$ and 
$B_d \rightarrow \phi K_S$, and the direct CP asymmetry in 
$B\rightarrow X_s \gamma$ {\it etc.}, and confront the measured data with
the SM predictions, in order to confirm the Kobayashi-Maskawa paradigm
or discover indirect new physics effects.

\bigskip
\acknowledgments

We are grateful to Intae Yu for discussions on the experimental data from
D\O\ and CDF\@.
PK is supported in part by KOSEF through the SRC program at
CHEP, Kyungpook National University.
JhP acknowledges Research Grants funded jointly by the Italian
Ministero dell'Istruzione, dell'Universit\`{a} e della Ricerca (MIUR),
by the University of Padova and
by the Istituto Nazionale di Fisica Nucleare (INFN) within the
\textit{Astroparticle Physics Project} and the FA51 INFN Research Project.
He was also supported in part by the European Community Research
Training Network UniverseNet under contract MRTN-CT-2006-035863.

\end{document}